\documentstyle[12pt,epsfig]{article}
\def\be{\begin{equation}}
\def\ee{\end{equation}}

\def\a{\alpha}

\def\e{\varepsilon}

\def\lm{L_{\rm min}}
\def\s{\sigma}

\begin{document}

\title{Metal-insulator transition in three dimensional
Anderson model: scaling of higher Lyapunov exponents}

\author{P. Marko\v s\\
Institute of Physics, Slovak Academy of Sciences,\\
D\'ubravsk\'a cesta 9, 842 28 Bratislava, Slovakia\\
e-mail address:  markos@savba.sk}

\maketitle

\begin{abstract}
Numerical studies of the Anderson transition are based 
on  finite-size scaling analysis of the smallest positive
Lyapunov exponent. We prove numerically that the same scaling
holds also for higher Lyapunov exponents. 
This scaling supports the hypothesis of the one-parameter 
scaling of the conductance distribution. 
From  collected numerical data 
for quasi one dimensional systems up to  system size $24^2\times\infty$
 we found the critical disorder 
$16.50\le W_c\le 16.53$ and the critical exponent
$1.50\le\nu\le 1.54$. Finite-size effects and the role of  irrelevant
scaling parameters are discussed.
\end{abstract}

\medskip


\newpage

The main problem of the theory of Anderson transition is to prove
that there is only one relevant parameter which controls the
behavior of all quantities of interest in the neighborhood of the
critical point. 
An excellent example of such  a  quantity 
is the smallest positive Lyapunov exponent (LE) $z_1$
which follows the one-parametric scaling relation
\cite{MacK}
\be\label{1}
z_1(L,W)=f(L/\xi(W))\footnote{Instead of $z_1$, the inverse quantity
$\Lambda=2/z_1$ is commonly used. 
The present discussion is identical
for both quantities, but  $z$'s seem to be more natural variables
for our purposes.}.
\ee
In (\ref{1}), $W$ is the disorder, $L$ defines the width of the
quasi-one dimensional system $L\times L\times L_z$
and $\xi(W)$ is the universal scaling parameter. In  Q1D
geometry, Lyapunov exponents $z_i$ are defined through eigenvalues
$t_i$ of the transfer matrix $T^\dag T$ 
as $z_i=2{L\over L_z}\log t_i$. In the limit $L_z>>L$, all $z$s are
self-averaged quantities 
\cite{oseledec}.  
Relation (\ref{1}) determines the disorder  and the system size 
dependence of $z_1$ in the neighborhood of the critical point $W_c$ and
enables us to determine $W_c$ and 
critical exponents for 
conductance ($s$, $W<W_c$) and for  localization length 
($\nu$, $W>W_c$)
\cite{MacK,MacKa,MacKinnon,SO,bulka,henneke}

In the pioneering work
\cite{MacK}, critical parameters were found as
 $W_c\approx 16.5\pm 0.5$ for the box distribution of the
disorder energies, and  $s=\nu\approx 1.5\pm 0.1$.
This result  was later confirmed  by more accurate
numerical studies
\cite{MacKinnon,OS,SO}, and also by analysis of the level statistics
\cite{shkl}.
Calculations performed for different microscopic models confirmed the
universality of exponent $\nu$ within a given universality class
\cite{bulka}.

To complete the 
proof of the universality of the metal-insulator transition,
the one parameter scaling should be found also for more realistic
variables, such as the conductance $g$
\cite{AALR,shapiro}. 
This must be done   for the cubic
samples. Here, owing  to the absence of self-averaging,
it is necessary to test the universal scaling of the
whole distribution $P(g)$. 
It is  unrealistic to perform such analysis with the numerical
accuracy comparable to that achieved from Q1D studies. Therefore,
previous studies of $P(g)$ concentrated  only  on  estimation 
of the conductance distribution at the critical point
\cite{pm,el,OS,SO,prl} and in the metallic and localized regime
\cite{pm}.

The aim of 
this  work  is to support the idea of the  one-parametric
scaling of the conductance and of its distribution. 
Instead of the study $g$, we 
prove numerically that the higher Lyapunov exponents
$z_2, z_3,\dots$ follow 
the same scaling behavior as the first one
in the Q1D systems.
Common scaling proves  that the spectrum of  the transfer matrix
in the Q1D limit is  determined only by one parameter. 
Strong correlation of $z$s gives
also the serious basis for the generalization of the random - matrix theory
to the description of the critical region
\cite{jpcm,jpa,mut}.

Although we  deal only with Q1D geometry, it is reasonable 
to suppose that the observed  
correlation survives also for the cubic geometry.  
Then the relation between $g$ and $z$s, 
$g=\sum_i\cosh^{-2}z_i/2$, \cite{pichardnato}
assures that   $g$  also follows  the one-parametric scaling.

Collected numerical data also provide us  with a very accurate estimation
of the critical disorder $W_c$ and the critical exponent $\nu$.
It is the first time that numerical data  for system size
$L>16$ has been collected and analyzed. Our data for large $L$ enable us to
check the finite-size corrections to scaling proposed in
\cite{SO}.

\medskip

The scaling behavior of higher LEs was originally  studied 
in the Henneke's PhD Thesis 
\cite{henneke}. 
Due  to the insufficient
accuracy of his data and small system size, no acceptable proof of the common
scaling was found.  
The first indication of the common scaling was shown in
\cite{pmnato}
and generalized to the neighborhood of the critical point in
\cite{jpcm,jpa}.
The common behavior of higher LEs, 
$z_i\sim i$ is well known in the metallic regime; it was  already used in
\cite{MacK} to explain the physical meaning of the scaling parameter $\xi(W)$,  and confirmed by random-matrix studies
\cite{pichardnato}.
In the localized regime, $z$s follow the relation $z_i(W,L)=z_1(W,L)+\Delta_i$
with $W-$ and $L- $ independent constants $\Delta_i$
\cite{jpcm}.

\medskip

For the Q1D systems $L^2\times L_z$
we calculated all LEs for cca 21 different values of disorder,
$16\le W\le 17$. $L$ grows from $L=4$ up to 24.
For the smallest $L$, the  relative accuracy of the first LE $z_1(W,L)$,
$\e_1=\sqrt{{\rm var} z_1}/z_1(W,L)$  was 
0.05\% while 
$\e_1$ was  only 1\% for $L=21,22$ and 24, being 0.5\% for $L=16,18$.
The accuracy of higher LE is much better;
in particular, $\e_2\approx \e_1/2$ and $\e_9\approx 0.17\e_1$ for all system
size.

The interval of the disorder is narrow enough to approximate the $W$ dependence of $z$'s by the linear fit
\be\label{twopar}
z_j(W,L)=z_j^{(0)}(L)+Wz_j^{(1)}(L), \quad j=1,2,\ldots
\ee
Small differences between 
fits containing higher powers of $W$  and  (\ref{twopar}) appear 
only for $L>18$ and
even then they do  not exceed the numerical inaccuracy of the raw data.
The typical $W$ - dependence of our data is presented in Fig. 1 for $z_2$.

\medskip

The scaling behavior requires that \cite{progr}
\be\label{scal}
z_j(W,L)\approx z_{jc}+A\times (W-W_c)\times L^\a,\quad\quad \a=1/\nu.
\ee
Comparison of (\ref{twopar}) with (\ref{scal}) offers the simplest way
to estimate the critical exponent $\a$. In Fig. 2. we present
the $L$-dependence of $z_j^{(1)}$ for the first six LEs and for $z_9$. It confirms that
close to the critical point these LEs scale with the same exponent $\a$: 
\be\label{alpha}
\alpha\approx 0.655\pm0.010
\ee
which determines $\nu=1.526\pm 0.023$.
This estimation  is in very good agreement with the result of MacKinnon
\cite{MacKinnon} and  differs slightly  from 
\cite{OS,SO}.

\medskip

Figs. 1 and 2 also show also the  important influence
of the finite-size effects (FSE) in the
present analysis. Evidently, the small $L$ data are  of no use in the
analysis of higher LEs. We found that the  numerical data  for  $z_j$ 
could be used only when 
\be\label{cond}
L>z_j. 
\ee
It is easy to understand. 
If $z_j> L$ then the $j$th channel is rather "localized" than critical
on this length scale. Therefore  
only a small part of the spectra which fulfills the relation (\ref{cond}) 
follows the scaling behavior. The rest of the spectrum depends on $L$
even at the critical point. This conclusion is supported also by the 
analysis of the density $\rho(z)$ of all LEs for the cubic samples 
\cite{loc99}. 
At the critical point, the number of system-size independent LEs
grows as $\sim L$ when $L\to\infty$
\cite{jpcm}.
As $z_1\approx 3.4$, the above mentioned 
effect does  not influence  the analysis
of the first LE $z_1$.
Nevertheless, other FSE must be taken into account
in the scaling analysis  of $z_1$
\cite{SO,ann}.

\medskip

More reliable  estimation of the exponent $\a$ (\ref{alpha})
and of the critical disorder $W_c$ is given by the position of the 
minimum of the function
\be\label{minimum}
F(W_c,\a,\ldots)=\frac{1}{N}\sum_{W,L}\frac{1}{\s_j^2(W,L)}\big[z_j(W,L)-
z_j^{\rm fit}(W,L)\big]^2.
\ee
In (\ref{minimum}),  $N=\sum_{W,L}$  is the number of points, 
and \dots  stays for all other fitting parameters. 

The natural choice of the fitting function $z_j^{\rm fit}$ in (\ref{minimum}) is the rhs of
(\ref{twopar}). 
None FSE are explicitly included in (\ref{twopar}). 
Nevertheless, it still enables us to 
test the sensitivity of the critical parameters to the  size 
of the analyzed systems. To do so, we considered 
different sets of input data $z_j(L,W)$ with  the restriction 
$\lm\le L\le L_{\rm max}$ ($\lm\le 12$).
Then, the  
$\lm$- and $L_{\rm max}$- dependence of $W_c$ and $\alpha$ was analyzed.
While the influence of the choice of  $L_{\rm max}$ is, as supposed, negligible,both $W_c$ and $z_{jc}$ increase with $\lm$. 
We found the $\lm$ - independent results only for the two 
smallest LEs $z_1$ and $z_2$.
For the higher LEs, critical parameters do not reach their limiting values
even for $\lm=12$. 
In difference to $W_c$,  the estimation of the critical exponent $\a$ does
not depend on the choice of interval of $L$. Obtained data are in 
good agreement with estimation (\ref{alpha}) for all LEs under consideration. 

The weak $\lm$ - sensitivity of the critical exponent  agrees with an assumption that FSE
influence primarily  the $W$-independent part of $z_j$
\cite{MacKinnon}.
Fig 1. offers a simple interpretation of this result: to eliminate 
FSE one has to  shift  each line by the disorder-independent 
constant $B(L)$ which should be added to the rhs. of (\ref{twopar}). 
The proper
choice of  $B(L)$, assures that all lines  cross at  the same point as it is
proposed by the scaling theory. Finite size corrections to the line slope 
are  only of the "higher order".

Slevin and Ohtsuki
\cite{SO}
fitted $z_1(W,L)$ (resp. its inverse $z_1^{-1}$) to the  more general function
\be\label{genpar}
z_1^{\rm fit}(L,W)=z_{jc}+\sum_{n=0}^{N_x}\sum_{m=0}^{N_y}A_{nm}x^ny^m
\ee
with $N_x=3$, $N_y=1$. In (\ref{genpar}),
$x=(w+b_1w^2+b_2w^3)L^\alpha$, 
$w=W-W_c$ and
$y=L^\beta$ with $\beta<0$. Exponent $\beta$ represents 
the second critical (irrelevant) scaling  exponent.
We applied this function to our data with restriction (\ref{cond})
and with $b_1=b_2=0$, $n+m\le 1$. Then
\be\label{threepar}
z_j^{\rm fit}(L,W)=z_{jc}+A\times(W-W_c)L^\a+BL^\beta.
\ee
We have checked that
more sophisticated fits do not provide us with any reasonable improvement of
the accuracy of critical parameters. 

To test the quality
of the fit (\ref{threepar}), we again studied the 
sensitivity of our results to a
change of the input  data. 
Evidently, for  large enough $\lm$ the role of
 the irrelevant scaling exponent is negligible. 
The finite size effects become small and  difficult to  measure.
Value of the 
irrelevant critical exponent $\beta$ obtained from fitting function
(\ref{threepar})  decreases
to  $\approx -20$ for large $\lm$.

For small values of $\lm$, however, 
the three-parametric fit (\ref{threepar})
still  does not provide us with the 
$\lm$-independent estimation of critical parameters.
We averaged therefore the  values
of $W_c$ and $\alpha$ obtained from  
various choices of $\lm$.

Table 1. presents our results for the first five LEs 
obtained from  fits (\ref{twopar}) and (\ref{threepar}).
On the basis of the presented data we estimate 
\be\label{result}
16.50\le W_c\le 16.53\quad{\rm and}\quad 1.50\le\nu\le 1.54
\ee
These values are in a very good agreement with 
\cite{MacKinnon}.

Our results (\ref{result}) differ from that obtained by many parametric
fitting procedure in Ref.
\cite{SO} (Table 1.).
None of the analyzed  statistical ensemble  provides  
us with such high value of $\nu$.  
This  discrepancy is probably caused by
different input data. 
Contrary  to previous treatments
\cite{MacKinnon,SO},
we  collected  data for large system size in  order to simplify
the fitting function. The  main
shortcoming of this strategy is a lower accuracy of our data for $z_1$.
On the other hand, 
the fact that the results obtained  from  the many parametric fitting 
procedure  depend on $\lm$ 
indicates that the fitting function (\ref{threepar}) is still
insufficient to reflect completely the corrections to scaling. 
The only way to obtain a more accurate  estimation of the critical
parameters is to collect more exact numerical data for
large system size. 

\medskip

To conclude, we have collected numerical data for the quasi one dimensional
Anderson model up to  system size $L=24$. Our data prove that
higher Lyapunov exponents 
of the transfer matrix follow the one-parametric scaling
law.  The  critical exponent 
$\nu$  coincides with that calculated from the
scaling treatment of the  smallest LE. The scaling holds
only for Lyapunov exponents which are smaller than the system size considered.

The common scaling enables us to express  all 
relevant LEs as the unique function 
of the first one. Evidently, the same holds also for any
function of $z$s. This indicates the validity of the scaling theory
for the conductance.  
However, our analysis was restricted to the
quasi-one dimensional systems.  Rigorous proof  
of the one-parametric scaling of the conductance requires  repeating
the performed  scaling analysis for the cubic samples, 
where no self-averaging of $z$s and of $g$ takes place. 

We show  for the first time, that the numerical data for higher LE
could be  used for calculation of critical parameters of the metal-
insulator transition. The numerical accuracy 
of higher  LE  is much better than that of $z_1$. 
The price we  pay for
it is a  stronger influence of the finite-size effects which causes 
that the data
obtained for small system size cannot be used for the scaling analysis.
The best compromise between the accuracy and FSE offer
data for the second LE $z_2$. 
We discussed the methods of elimination   of the finite size effects
and estimated the  critical disorder and the 
critical exponent $\nu$ by relation (\ref{result}).

\medskip

\noindent{\bf  Acknowledgment} This work has been supported by Slovak Grant Agency,
Grant n. 2/4109/98. Numerical data have been partially 
collected using the computer
Origin 2000 in the Computer Center of the Slovak Academy of Sciences.

%\end{document}
\newpage
\newpage

{\footnotesize
\begin{table}[h]
\begin{tabular}{|l|c|c|l|l|l|l|l|}
\hline
$j$ &    $\lm$ & $L_{\rm max}$ &  $W_c$ &$z_{jc}$  &    $\a$  & $\nu$ &   $\beta$  \\
\hline
1   & 4-5 & 24   &   16.515  & 3.46 &  0.644 & 1.55  & -3.5 \\
$1^*$ & 8-12  & 20-24   &  $16.505~(10)$ & $3.451~(07)$ &   $0.681~(15)$ & 1.470~(30)  &   $-$      \\ 
2   &  5-10  & 24 & $16.527~(02)$  & $5.588~(02)$  & $0.654~(08)$ &   1.529~(18)  &-3.2~(6)   \\
$2^*$ & 10-12  & 22-24 &  $16.500~(07)$ &$ 5.500~(07)$ &  $0.659~(05)$ &   1.517~(11)  &  $-$   \\
3   &    9  & 24  &16.508 & 7.167  & 0.647 &  1.545  &-6.0    \\
4   &  8-10  & 24 & $16.504~(02)$  & $8.422~(05)$  & $0.663~(04)$ & 1.509~(9)  &-3.7~(2) \\
5   &  9-12  & 24 & $16.517~(16)$&  $9.560~(30)$ & $0.661~(06)$ &   1.513~(14) &-3.3~(8)   \\
\hline
$1^{\cite{MacKinnon}}$ &4 &12 & $16.500~(50)$ & & & $1.515~(33)$   & \\
$1^{\cite{SO}}$       &6 & 12 & $16.448~(14)$ & & & $1.59~(3)$    &  \\
$1^{\cite{OS}}$       &4 & 14 & $16.540~(10)$ & & & $1.57~(2)$   & -2.8~(5)   \\
$1^{*\cite{OS}}$      &8 & 14 & $16.514~(07)$ & & & $1.58~(5)$   & -   \\
\hline
\end{tabular}
\caption{Critical disorder $W_c$ and critical exponent $\nu$  as have been found from numerical data for 
the $j$th LE for the three-parametric and two-parametric fits ($^*$) and their
comparison with other results. Number of analyzed points is $\approx 21\times
(L_{\rm max}-\lm)$. The minimum of $F$ was found $\le 1.05$ for all analyzed
sets (with exception of $z_3$, where it was 1.09).}
\end{table}
}
\newpage
\newpage

\begin{figure}[h]
\epsfig{file=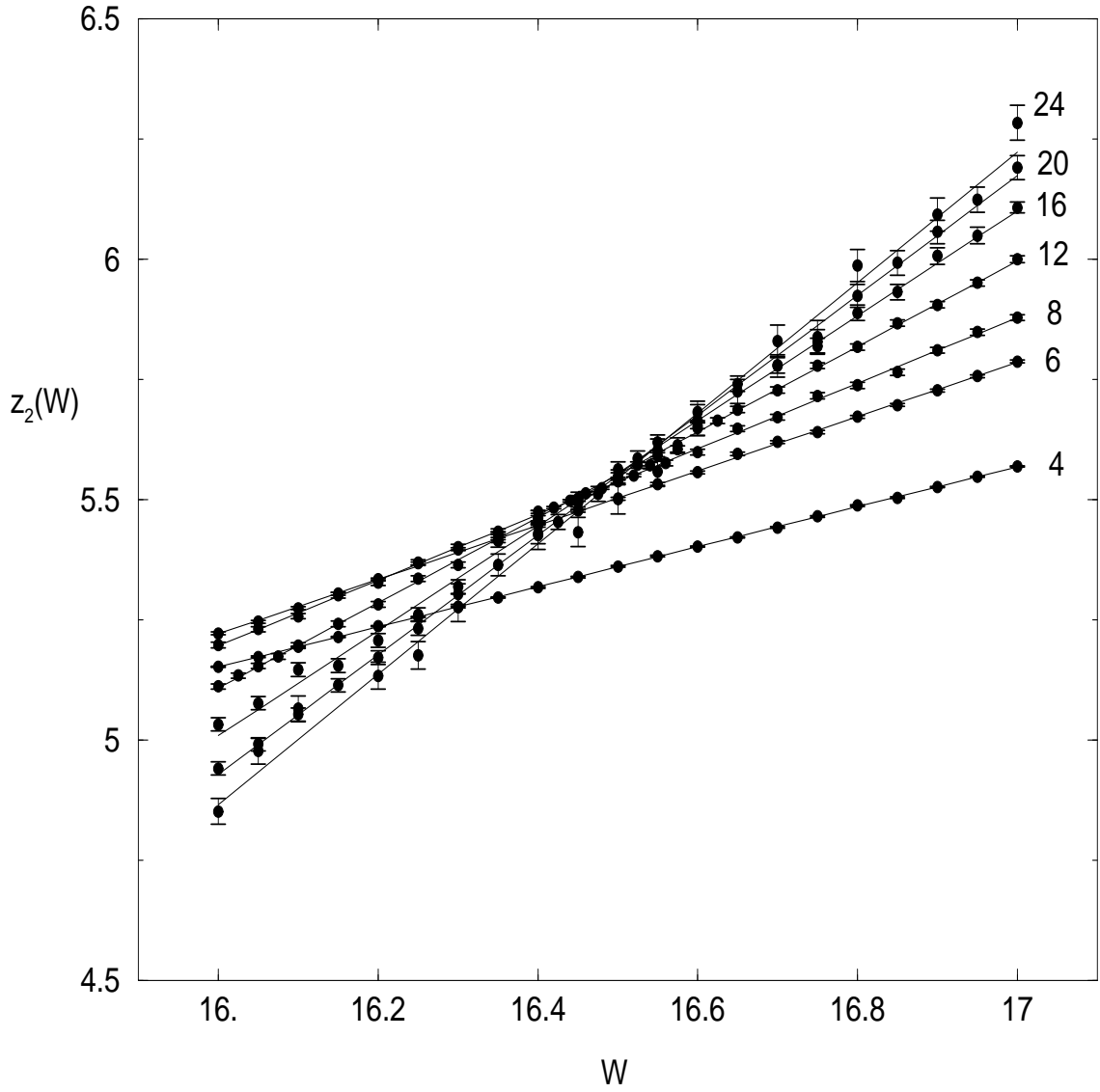,width=15cm,height=15cm,angle=-90}
\caption{The $W$-dependence of the second LE $z_2$ for different
system size.}
\end{figure}

\newpage
\newpage

\begin{figure}[h]
\epsfig{file=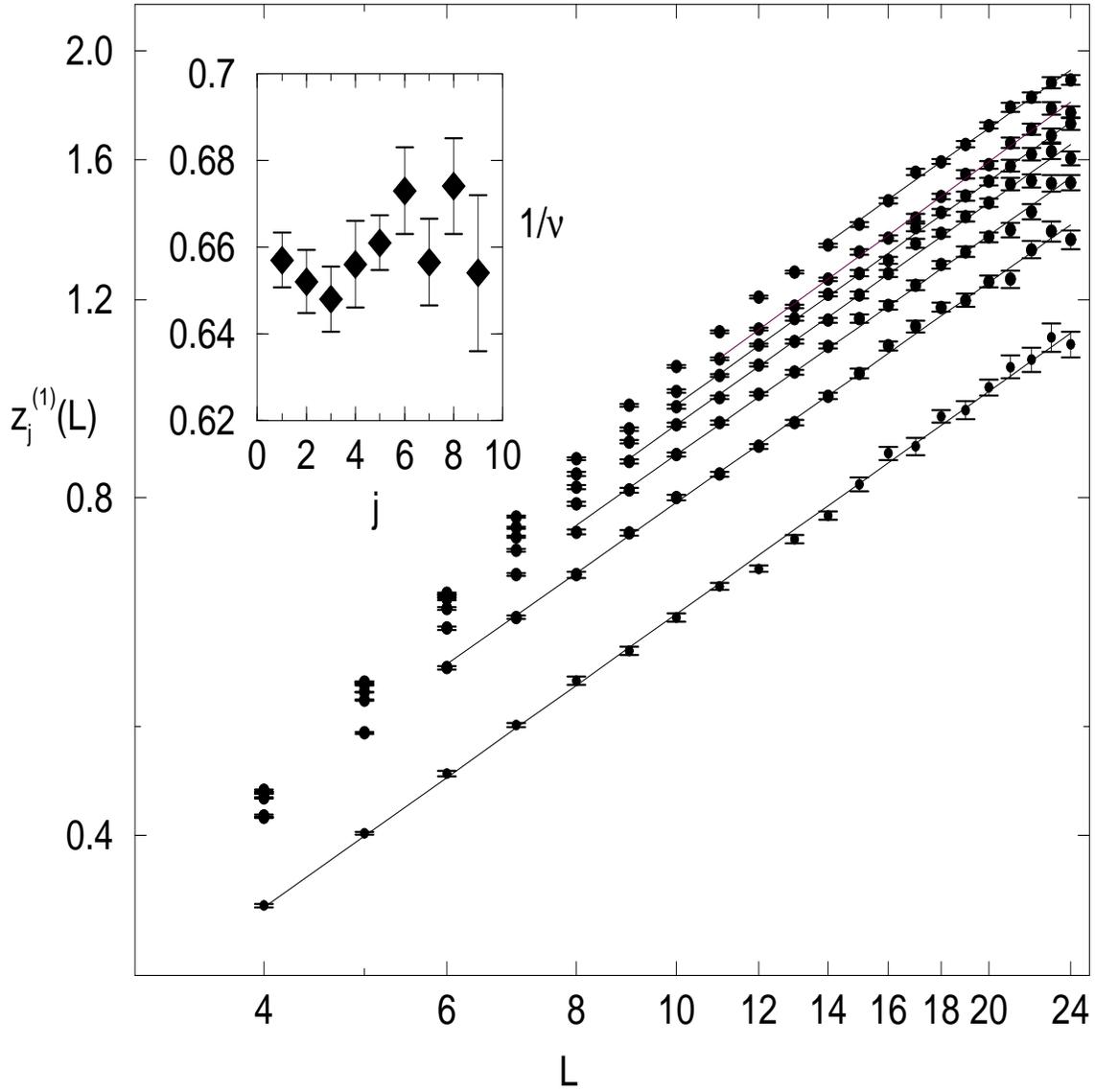,width=15cm,height=15cm}
\caption{The $L$-dependence of $z_j^{(1)}$ for the first six LE and for
$z_9$ (counted from below). The slope determines critical exponents as $\a=1/\nu$. Inset:
values of $1/\nu$ found from presented fits.}
\end{figure}

\end{document}